\documentclass[apj, twocolumn]{aastex63}
\usepackage{multirow,acronym}
\usepackage{natbib}
\usepackage{hyperref}
\usepackage{tikz}
\usepackage{pgfplots}
\pgfplotsset{compat=1.9}
\usepackage{booktabs}
\usepackage{amsmath}
\usepackage[T1]{fontenc}
\usepackage{graphicx}
\usepackage{bm}
\usepackage{subfigure}
\hypersetup{colorlinks=true, citecolor=blue, linkcolor=cyan, urlcolor=magenta}

\usepackage{epstopdf} 

\usepackage{lipsum}

\newcommand{\h}{$^{\rm h}$}
\newcommand{\m}{$^{\rm m}$}
\newcommand{\s}{$^{\rm s}$}

\newcommand{\TSS}{TianQin Research Center for Gravitational Physics and School of Physics and Astronomy, Sun Yat-sen University (Zhuhai Campus), Zhuhai 519082, China}
\newcommand{\SPA}{School of Physics and Astronomy, Sun Yat-sen University (Zhuhai Campus), Zhuhai 519082, China}

\begin{document}


\title{Following up the afterglow: strategy for X-ray observation triggered by gravitational wave events}

\author{Hui Tong}
\thanks{These two authors contributed equally}
\author{Mu-Xin Liu}
\thanks{These two authors contributed equally}
\affiliation{\SPA}

\author{Yi-Ming Hu}
\email{huyiming@mail.sysu.edu.cn}
\affiliation{\TSS}

\author{Man Leong Chan}
\affiliation{Department of Applied Physics, Fukuoka University, Nanakuma 8-19-1, Fukuoka 814-0180, Japan}

\author{Martin Hendry}
\affiliation{SUPA, School of Physics and Astronomy, University of Glasgow, Glasgow G12 8QQ, UK}

\author{Zhu Liu}
\author{Hui Sun}
\affiliation{Key Laboratory of Space Astronomy and Technology, National Astronomical Observatories, Chinese Academy of Sciences, Beijing 100101, China}


\acrodef{GW}{gravitational wave}
\acrodef{EoS}{equation of state}
\acrodef{EM}{electromagnetic}
\acrodef{FOV}{field of view}


\begin{abstract}
  The multi-messenger observation of compact binary coalescence promises great scientific treasure.
  However, a synthetic observation from both gravitational wave and electromagnetic channels remains challenging. 
  Relying on the day-to-week long macronova emission, GW170817 remains the only event with successful electromagnetic followup. 
  In this manuscript, we explore the possibility of using the early stage X-ray afterglow to search for the electromagnetic counterpart of gravitational wave events.
  Two algorithms, the \emph {sequential observation} and the \emph {local optimization} are considered and applied to three simulated events.
  We consider the proposed \emph{Einstein Probe} as a candidate X-ray telescope.
  Benefiting from the large field of view and high sensitivity, we find that the sequential observation algorithm not only is easy to implement, but also promises a good chance of actual detection. 
\end{abstract}


\section{Introduction}

The successful operation of ground-based \ac{GW} detectors like LIGO and Virgo has marked the beginning of a new era of \ac{GW} astronomy \cite{abbott2016observation,2016ApJ...818L..22A,2016PhRvL.116f1102A,2016PhRvL.116m1103A,2016PhRvD..93l2003A,2016PhRvL.116x1102A,2016PhRvL.116v1101A,2016ApJ...833L...1A,2016PhRvD..93l2004A,2016CQGra..33m4001A,2016PhRvL.116m1102A,2017PhRvD..95f2003A}. 
Throughout the first to the third observation run of LIGO and Virgo, a series of detection of compact binary coalescence has been made, containing a large variety of features like mass, mass ratio, spin, and distance, etc.\cite{2016PhRvX...6d1015A,2016PhRvL.116x1103A,2017PhRvL.118v1101A,2017ApJ...851L..35A,2017PhRvL.119n1101A,2017PhRvL.119p1101A,2017ApJ...848L..12A,2017ApJ...848L..13A,2017Natur.551...85A,Abbott:2020uma, 2020arXiv200408342T}.
Following the excellent traditions of observational astronomy, where the opening of each new observation channel promises great scientific return, the operation of \ac{GW} observatories greatly deepens our understanding to the most dense objects in the Universe.

Apart from the window of \ac{GW}, other observational channels like \ac{EM}, neutrinos and cosmic rays, often referred to as ``messengers'', can also be used to study the properties of the underlying sources.
The joint force of multi-messenger observation of a specific systems opens an exciting potential for astronomers.
The simultaneous optical and neutrino observations of SN1987A set limits to properties of neutrinos \cite{1989ARA&A..27..629A}.
In 2017, the neutrino detector IceCube detected a high-energy neutrino, IceCube-170922A, hinted blazars as possible origin of the high energy neutrinos \cite{2018Sci...361.1378I}.
The reported flash 0.4 seconds after GW150914 claimed by Fermi GBM triggered huge interests among astronomers \citep[e.g.][]{2016ApJ...826L...6C,Li:2016zjc}
In fact, even prior to the first direct \ac{GW} event, the non-detection of \ac{GW} signals could already be used to put constraints on certain sources under the multi-messenger framework, like limiting the origin of GRB 070201 \cite{2008ApJ...681.1419A}, or deduce the property of \ac{EoS} \cite{2017ApJ...844L..22L}.

In 17 August, 2017, the \ac{GW} signal from a binary neutron star merger was observed as GW170817 \cite{2017PhRvL.119p1101A}, and a GRB event GRB170817A was observed simultaneously around the same location \cite{2017ApJ...848L..13A}.
Such simultaneous observations of both \ac{GW} and \ac{EM} waves sparked huge interests among astronomers, leading to a series of scientific discoveries, from confirming the link between short gamma ray burst and binary neutron star merger, to revealing its role in producing heavy elements throughout the Universe \cite{2017ApJ...848L..13A}.

However, notice that due to the exceptional nature of GRB170817A, which is the closest short GRB detected by human, it is the only event with multi-messenger observations from both \ac{GW} and \ac{EM} channels.
It is also by far the weakest short GRB in terms of luminosity \cite{2017ApJ...848L..13A}, 
No detection of \ac{EM} counterpart of the next neutron star binary merger event GW190425 was made \cite{Abbott:2020uma}.
The status quo of multi-messenger observation triggered by \ac{GW} detections reflect its intrinsic difficulty.

Although GRBs are powerful sources of light, they are highly beamed.
Only those who locate in a small solid angle close to its collimated jet could easily detect it \cite{1999ApJ...519L..17S}.
Detections of gamma rays often come with large uncertainties in their sky locations, therefore, even if the gamma ray burst is detected, the pinpointing of the location or identifying the host galaxy is still a needle-in-a-haystack search.
The successful identification of host galaxy for GRB170817A relies on the much later stage emission called kilonova or macronova, where infrared emission powered by decay of radioactive materials produced in the so-called \emph{r-}process after the binary neutron star merger \cite{1998ApJ...507L..59L,2013ApJ...775...18B}.
Compared with the prompt emission of GRB, macronovae has the advantage of much wider viewing angle \cite{2012ApJ...746...48M}, however, it suffers from lower luminosity.

In this work, we explore the X-ray afterglow stage of emission, to assist the rapid localization of a binary neutron star merger.
The X-ray afterglow is expected to happen sooner compared with the macronova, enabling the observation of earlier stage phenomenon \cite{1998ApJ...497L..17S}.
On the other hand, X-ray afterglow can be observed at a larger viewing angle compared with the highly beamed gamma ray emission.
More importantly, the duration for gamma ray burst prompt emission is too short to perform target of opportunity observation, while X-ray afterglow can last long enough to perform maneuver triggered by \ac{GW} alerts. We aim to study that under the assumption that a trigger from \ac{GW} observatories has been issued, and no short GRB has been observed, what observation strategy should X-ray telescopes adopts, so that one can increase the chance of observing the \ac{EM} counterpart and pinpoint its sky location.

This paper is organized as follows. 
Section \ref{sec:method} describes the statistical framework we adopt.
Section \ref{sec:algorithm} illustrates the two algorithms we uses for the observation strategy.
We show the results in section \ref{sec:results}, and discuss future work and summary in section \ref{sec:conclusion}.

\section{Statistical framework} \label{sec:method}
In order to optimize the observation strategy, we need to first determine the statistical framework.
Since the luminosity of X-ray afterglows changes rapidly, we define the detection as when multiple observations reveals obvious luminosity difference.

Throughout the work, we make certain assumptions. 
For example, we assume that the sky position and distance are independent of each other, so that their joint probability distribution is simply the multiplication of each distribution.
We also assume that the telescopes can point to any direction, ignoring the potential influence from the Sun, the Moon, and the Earth \cite{Chan:2015bma}.

We use $D_{\rm ag}$ to denote the successful detection of an afterglow. In order to confirm the existence of afterglow, one need to observe the change of luminosity, so $D_{\rm ag}$ is only defined when multiple observations are finished.
The probability of $D_{\rm ag}$ is defined as when the inferred flux has obvious difference, or $\Delta f > 0$.
We note that the probability of detection $P(D_{\rm ag})$ depends on the \ac{FOV} $ \omega $, the observed sky locations $ (\alpha,\delta) $, and the corresponding exposure time of the multiple observations $ \tau_1$ and $ \tau_2$. The posterior probability of successful detection can then be given as the probability of one first measure the flux for the first exposure, then observe an obvious change in flux between the two exposures:
\begin{equation}\label{1}
    \begin{aligned}
        &P(D_{\rm ag}|\omega,\tau_1,\tau_2,I)= \\
        &P(N>N^*|\omega,\alpha,\delta,\tau_1,I) \times P(\Delta f > 0|\tau_1,\tau_2,I)
    \end{aligned}  
\end{equation}

Here, $I$ is prior information that includes the parameters of the selected telescope, such as its photon collecting area $A$ to name one. The threshold count $N^*$ is the criterion for detection determined by the expected signal-to-noise ratio (SNR), the background noise, and the sensitivity of the selected telescope. We adopt the definition of SNR 
\begin{equation}\label{2}
    \begin{aligned}
        {\rm SNR}= \frac{N_{\rm signal}}{\sqrt{N_{\rm noise}}}
    \end{aligned}  
\end{equation}
hence, $N^*$ could be expressed as
\begin{equation}\label{3}
    \begin{aligned}
        N^*= {\rm SNR}\times\sqrt{N_{\rm noise}}
    \end{aligned}  
\end{equation}
The flux received by the telescope depends on multiple factors.
The GW event is localised by \ac{GW} detectors with uncertainties, and one can assess how likely a certain area contains the \ac{GW} source.
With the knowledge of luminosity and the distance $R$, one can estimate the distribution of the expected flux, which can be later translated into the distribution of detected photon numbers. 
Then the first part of Equation (\ref{1}) can be expanded as
\begin{equation}\label{4}
    \begin{aligned}
        P&(N>N^*|\omega,\tau_1,I)=\int^\infty_{N^*}dN\int df\int dR\int _\omega d\alpha d\delta\\&\times p(N|f,\tau_1,I)p(f|I,R)p(\alpha,\delta,R|I).
    \end{aligned}
\end{equation}
The quantity $P(N|f,\tau_1,I)$ is the probability of received photons, given the flux f of the source and observation time $\tau_1$, which is described by a Poisson distribution. Since we assume that the prior distribution on the distance to the target afterglow is statistically independent of the prior distribution on its sky location, Equation (\ref{4}) can be written as 
\begin{equation}\label{5}
    P(N>N^*|\omega,\tau_1,I)=P_{gw}(\omega) \times P_{\rm ag}(\tau_1)
\end{equation}
where
\begin{equation}\label{6}
    P_{gw}(\omega)=\int_\omega p(\alpha,\delta|I)d\alpha d\delta\\         
\end{equation}

\begin{equation}\label{7}
    \begin{aligned}
        P_{\rm ag}(&\tau_1)=\int df \int dR \int^\infty_{N^*}dN\\
        &\times p(N|f,\tau_1,I)p(f|I,R)p(R|I)\\
        &=\int df\int^\infty_{N^*}dN p(N|f,\tau_1,I) \\
        &\times \int dR  p(f|I,R)p(R|I)
    \end{aligned}  
\end{equation}

The first part of Equation (\ref{1}) only considers single observation. As for the second part of Equation (\ref{1}), $\Delta f$ is the different flux of the multiple observations(called $f_1$ and $f_2$) at different moments. And the $f_1$ and $f_2$ can be approximated by a distribution depending on the prior known flux $f'$, which is based on the afterglow light curve model. The equation of this distribution can be written as
\begin{equation}\label{8}
    \begin{aligned}
        P(f_0|&f',\tau)=\sum_{N=0}^\infty P(N|f',\tau) \times P(f_0|N,\tau)\\
        &=\sum_{N=0}^\infty P(N|f',\tau) \times \dfrac{P(N|f_0,\tau)P(f_0)}{\int_0^\infty P(N|f,\tau)P(f)df}
    \end{aligned}
\end{equation} 
Here $P(f)$ is the prior probability of the flux emitted by afterglow.
$P(N|f,\tau)$ can be approximated by Poisson distribution. For convenience of description, we refer to the probability in Equation (\ref{7}) as $P_1$,  
\begin{equation}\label{10}
    \begin{aligned}
        P_1=P_{\rm ag}(&\tau_1)=\int df \int dR \int^\infty_{N^*}dN\\
        &\times p(N|f,\tau_1,I)p(f|I,R)p(R|I)\\
        &=\int df\int^\infty_{N^*}dN p(N|f,\tau_1,I) \\
        &\times \int dR  p(f|I,R)p(R|I)
    \end{aligned}  
\end{equation}
And the second part in Equation (\ref{1}) as $P_2$,
\begin{equation}\label{11}
    \begin{aligned}
        P_2=P(\Delta f > 0|\tau_1,\tau_2,I)
    \end{aligned}  
\end{equation}
Thus the Equation (\ref{1}) could be written as
\begin{equation}\label{9}
    \begin{aligned}
        P(D_{\rm ag}|\omega,\tau_1,\tau_2,I)=P_{\rm gw}\times P_{1}\times P_{2}
    \end{aligned}
\end{equation}


In the Equation (\ref{1}), we only consider the probability of detecting the afterglow of one observation field. In actual observations, the more likely scenario is that multiple fields would be observed.
The total number of fields can be estimated as follows. Assuming that the GW sky localization error region covers S $deg^2$ and the size of the telescope FOV is $\omega$ $deg^2$, the maximum number of fields n can be estimated as n $\lesssim$ S/$\omega$ in the case of small FOV \cite{Chan:2015bma}.  If a large FOV is considered, more fields than $n'=S/\omega$ may need to be observed due to the GW sky localization error region in the shape of strip. There will even overlap between the fields. In this article, we mainly consider the situation with a large FOV.

We will not set a constrained total observation time at first, but the observation time do have natural restraint. The luminosity will decrease and when the signal is so small, we can not do a valid observation to achieve the expected SNR any longer. In other words, $P_{1}$ will not increase at that time. We mark this time as $T_{threshold}$. When the time has exceeded $T_{threshold}$, we no longer consider doing the first time observation for new fields.

\section{Algorithm}\label{sec:algorithm}
\subsection{Tiling}


There are a lot of schemes for tiling and scheduling \cite{Coughlin:2018lta,Ghosh2017,Rana2019,Coughlin2019}. Here we use greedy algorithm to optimize the tiling of the observing fields \cite{Chan:2015bma}. Firstly, find out the number of HEALPix that can just meet the required confidence level (in our case we set it to 90\%) in order to simplify the calculation. Then, the observation field is divided according to the size of the FOV. Each time, the field with the maximum detection probability will be output and named in order from one to n. Hence their label indicates their rank in terms of enclosed GW probability.

We use the telescope, Einstein Probe (EP, \cite{yuan2015einstein}\footnote{\url{http://ep.nao.cas.cn}})'s Wide-field X-ray Telescope (WXT module \footnote{\url{http://ep.nao.cas.cn/epmission/epinstruments/201909/t20190916_516240.html}}), as an example in our work, and its FOV is about 3600 $deg^2$ \cite{yuan2015einstein}. Because of the large FOV, the observation field covers a large area around the center point, which means that the area of the observation field may exceed the range previously determined by the 90\% confidence level. With the increase in the number n of observation fields, almost all areas of the GW probability distribution would be covered. The sum of $P_{GW}$ of all observation fields may be greater than 90\%, and even reach 99\% in some events. The result would be shown in Section 4.

\subsection{Sequential Observation Algorithm}

Firstly, we consider a relatively simple algorithm, Sequential Observation Algorithm(SO). As shown in Equation (\ref{5}), the probability can be separated to two parts, depending on the pointing directions and observation time $\tau$ respectively. Then one principle of our observation strategy can be given. As we mentioned before, the label of fields indicates their rank in terms of enclosed GW probability. If we choose to observe a new field, the label of this field must be the smallest among all the fields that never be observed.

We intend to complete the first time observation for as many field as possible until the signal is too weak to achieve the expected SNR. Then, we start the second time observation in the same sequence of fields. As for the time allocation, we make a change in observation time by 1 second each time and then compare the probabilities gained by each choice (that is, observing for different fields). So, considering that there is a slew time interval when making the telescope point at the new field, we compare the increment of $P_1$ when observing more ($1+slew time$) seconds in the old field with the probability gained by observing 1 second in the new field. We continue observing the old field till the the increment of $P_1$ is smaller than the probability gained by observing a new field. To be fair to compare the difference in the number of photons observed between two observations, we simply adopt the same time distribution in the second time observation. 

We can easily find that such a simple algorithm make large $P_{1}$ for the most fields. We complete the first time observation as many fields as possible when the signal is strong enough. Although when we do the second time observation the signal is weak, what determines $P_{2}$ indeed is the difference between the two observation signals. That is, no matter how weak the signal is, as long as the intensity of the first observation and the second observation are relatively different, then we will get a large $P_{2}$. We can find that getting large $P_{1}$ and large $P_{2}$ do not actually conflict, so we believe that this simple but useful algorithm SO will perform well.
\subsection{Local Optimization Algorithm}

Here we consider a more complex algorithm, Local Optimization Algorithm(LO), as a comparison. The first principle mentioned before is still followed. And here is another principle that if we choose to observe a field the second time, the label of this field must be the smallest among all the fields that has been observed. Based on these two principles, we can compare the probability to optimize the time allocation the same as Sequential Observation. The difference from SO is that we can choose to observe the same field for the second time before we finish the first time observation for all fields. This algorithm will get the local optimal result.

\section{Results} \label{sec:results}

In this section, we present the performance of our two different algorithms by using the example telescope EP to the follow-up of a few simulated GW events. These events have different error region areas as well as shapes, leading to the distinction of the tiling.
We test our algorithms on a couple of simulated events from \cite{Singer_2014}. 
Notice that these data were generated assuming the O1 sensitivity, which has already been surpassed by current detectors.
The same event is expected to have higher SNR, thus smaller sky localisation error.
In order to better mimic the realistic scenario, we manually multiply the distance of every event by a factor of 2.
The actual injected distances for our chosen events range from 30 Mpc to 100Mpc. However, which we updated to the range of 60Mpc to 200Mpc.

\begin{figure*} \centering    
\subfigure[GRB afterglows observations, together with fitted line] {
 \label{fig:subfig:GRB afterglows model}     
\includegraphics[width=1.0\columnwidth]{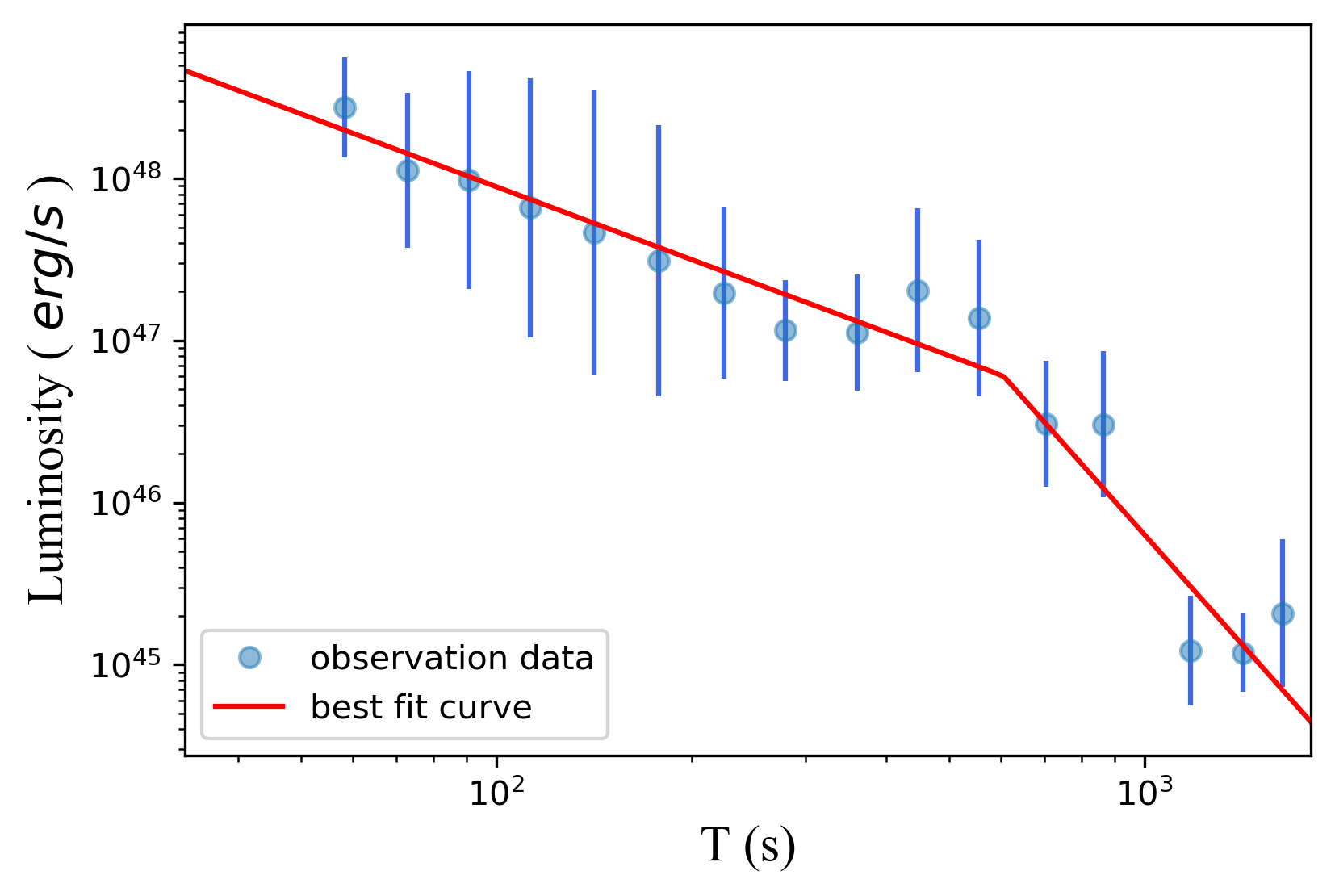}  
}     
\subfigure[X-ray transient observations, , together with fitted line] { 
\label{fig:subfig:X-ray Transient Model}     
\includegraphics[width=1.0\columnwidth]{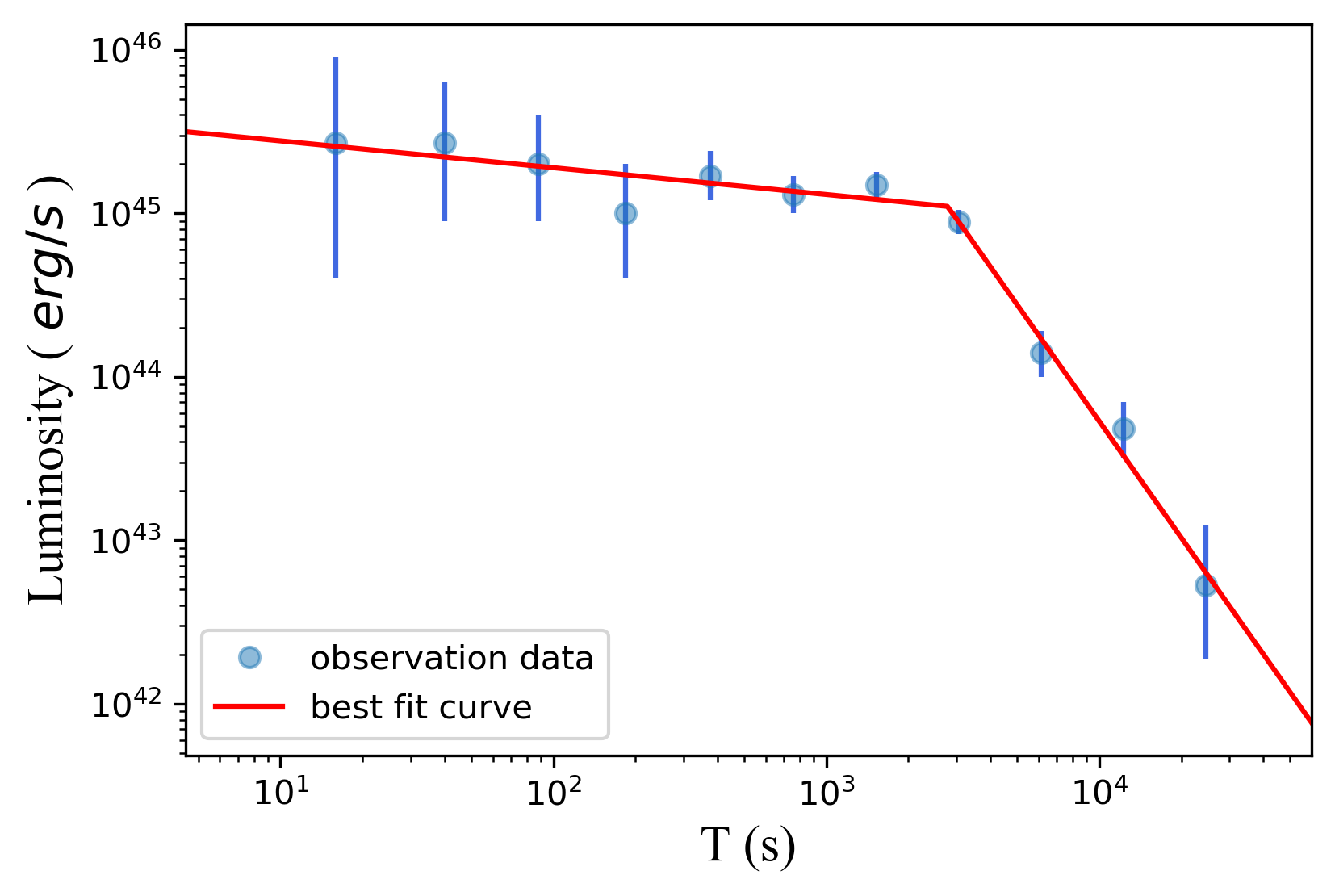}     
}    
\caption{Two different light curves is fitted from observations. The x-axis is the time after merger and the y-axis is the luminosity of the source. Notice that intrinsically the X-ray transient is much dimmer than the GRB afterglow.}     
\label{fig}     
\end{figure*}

We use two quite different light curve model respectively. The first light curve is shown in Figure \ref{fig:subfig:GRB afterglows model} , named as GRB Afterglows Model(GAM). It is fit based on the data from \footnote{\url{https://www.swift.ac.uk/xrt_curves/}}, more details about how the data were produced can be seen in \cite{Evans:2007na} and \cite{Evans:2008wp}. They are detected when compact binaries (NS-NS or NS-BH) merger and we are at the line of sight of the jet (or slightly off-axis). 
For later calculation, we fit the binned data with a two-step linear function in logarithmic space.
The second is shown in Figure \ref{fig:subfig:X-ray Transient Model} from \cite{Xue:2019nlf,Sun_2019}, called X-ray Transient Model (XTM) in our work. This kind of X-ray transient is produced when binary neutron stars merger and produce a highly-spinning massive neutron star, called magnetar (\cite{Dai2006,Gao2006}). These X-ray transients are more like isotropic and may not be related with GRBs (\cite{Zhang2013,Sun2017}). Both the X-ray afterglows and the X-ray transients can be observed if the merger leave a magnetar and we are close to the jet direction, while only X-ray transients are seen if the line of sight is off the jet axis. Generally both the two types of X-ray emissions can be regarded as the X-ray counterparts of gravitational events. So we work out both of them with our calculations.

\subsection{Tiling}


We obtained the optimized tiling of observing fields using the greedy algorithm approach for three typical events. Although EP's WXT has such a large FOV about 3600 $deg^2$, the error region in the shape of strip may result in a few observation fields needed to be observed in order to covering as much error region as possible. In our algorithms, we neglect the fields with small probability lower than about 1\%. So the required number and location of fields differ for different events.  The largest and smallest number of observation fields are 3 and 7. And there exists the difference of total $P_{gw}$ of all fields for different events but less than 3\%, which means we will covering error region more than 95\% for all events if we finish the detection of all fields except for the neglected ones.

\subsection{Time Allocation And Observation Sequence}

\begin{table*}

\begin{center}
\hspace*{-3.5cm}
\begin{tabular}{|c|c|c|c|c|c|c|c|c|} 
\hline
\multirow{2}*{Event ID}&\multirow{2}*{Algorithm}&\multicolumn{7}{|c|}{ allocated observation time (s)}\\
\cline{3-9}
&&Field1&Field2&Field3&Field4&Field5&Field6&Field7\\  
\hline
\multirow{3}*{10968}&SO&1&1&1&-&-&-&-\\  
\cline{2-9}  
&LO(First time)&1&1&1&-&-&-&-\\
\cline{2-9}
&LO(Second time)&18&1&53&-&-&-&-\\
\hline
\multirow{3}*{12715}&SO&1&1&1&1&1&1&-\\  
\cline{2-9}  
&LO(First time)&1&1&1&1&1&1&-\\
\cline{2-9}
&LO(Second time)&1&1&1&1&24&72&-\\
\hline
\multirow{3}*{14011}&SO&1&1&1&1&1&1&2\\  
\cline{2-9}  
&LO(First time)&1&1&1&1&1&4&13\\
\cline{2-9}
&LO(Second time)&1&1&1&7&220&83&75\\
\hline
\end{tabular}
\caption{Time allocation for the three example events}
\label{tab:Time allocation for three typical events}
\end{center}

\end{table*}

\begin{figure}{}
\tikzstyle{common} = [rectangle,rounded corners, minimum width=1.7cm,minimum height=1.3cm,text centered,text width=1.4cm,draw=black,fill=blue!30]
\tikzstyle{arrow} = [thick,->,>=stealth]
\tikzstyle{arrow2} = [thick,->,dashed,>=stealth]
\tikzstyle{arrow3} = [thick,-,dotted,>=stealth]
\begin{tikzpicture}[node distance=2cm]
\node (FieldOne1) [common] {Field1};
\node (FieldOne2) [common,below of=FieldOne1,yshift=-0.5cm] {Field1};
\node (FieldTwo1) [common,right of=FieldOne1,xshift=0.5cm] {Field2};
\node (FieldTwo2) [common,below of=FieldTwo1,yshift=-0.5cm] {Field2};
\node (FieldThree1) [common,right of=FieldTwo1,xshift=0.5cm] {Field3};
\node (FieldThree2) [common,below of=FieldThree1,yshift=-0.5cm] {Field3};
\node (FieldN1) [common,right of=FieldThree1,xshift=0.5cm] {Field N (if need)};
\node (FieldN2) [common,below of=FieldN1,yshift=-0.5cm] {Field N};
\draw [arrow] (FieldOne1) -- (FieldOne2);
\draw [arrow] (FieldOne2) -- (FieldTwo1);
\draw [arrow] (FieldTwo1) -- (FieldTwo2);
\draw [arrow] (FieldTwo2) -- (FieldThree1);
\draw [arrow] (FieldThree1) -- (FieldThree2);
\draw [arrow2] (FieldThree2) -- (FieldN1);
\draw [arrow] (FieldN1) -- (FieldN2);
\draw [arrow3] (FieldThree1) -- (FieldN1);
\draw [arrow3] (FieldThree2) -- (FieldN2);

\end{tikzpicture}
\caption{Illustration of observation sequence for Local Optimization algorithm.}
\label{fig:observation sequence for Algorithm Two}
\end{figure}
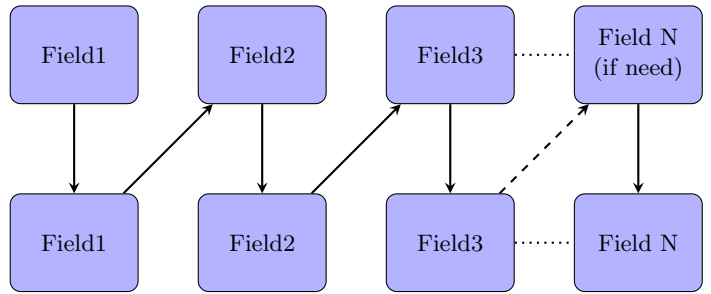



\begin{figure}{}
\tikzstyle{common} = [rectangle,rounded corners, minimum width=1.7cm,minimum height=1.3cm,text centered,text width=1.4cm,draw=black,fill=blue!30]
\tikzstyle{arrow} = [thick,->,>=stealth]
\tikzstyle{arrow2} = [thick,->,dashed,>=stealth]
\tikzstyle{arrow3} = [thick,-,dotted,>=stealth]
\begin{tikzpicture}[node distance=2cm]
\node (FieldOne1) [common] {Field1};
\node (FieldOne2) [common,below of=FieldOne1,yshift=-0.5cm] {Field1};
\node (FieldTwo1) [common,right of=FieldOne1,xshift=0.5cm] {Field2};
\node (FieldTwo2) [common,below of=FieldTwo1,yshift=-0.5cm] {Field2};
\node (FieldThree1) [common,right of=FieldTwo1,xshift=0.5cm] {Field3};
\node (FieldThree2) [common,below of=FieldThree1,yshift=-0.5cm] {Field3};
\node (FieldN1) [common,right of=FieldThree1,xshift=0.5cm] {Field N (if need)};
\node (FieldN2) [common,below of=FieldN1,yshift=-0.5cm] {Field N};

\draw [arrow] (FieldOne1) -- (FieldTwo1);
\draw [arrow] (FieldTwo1) -- (FieldThree1);
\draw [arrow2] (FieldThree1) -- (FieldN1);
\draw [arrow] (FieldN1) -- (FieldOne2);
\draw [arrow] (FieldOne2) -- (FieldTwo2);
\draw [arrow] (FieldTwo2) -- (FieldThree2);
\draw [arrow2] (FieldThree2) -- (FieldN2);

\end{tikzpicture}
\caption{Illustration of observation sequence for Sequential Observation algorithm.}
\label{fig:observation sequence for Algorithm one}
\end{figure}
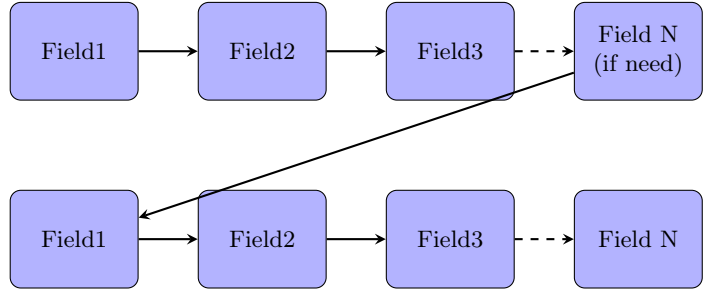
The work flow \ref{fig:observation sequence for Algorithm Two} shows the results of observation sequence for different algorithms used in four typical events, using GRB Afterglows Model. For Local Optimization, the strategy always requires two consecutive observations of the same field, which is determined by two factors. First, local optimization is to consider the effect of both $P_{gw}$ and $P_{em}$. Since the fields with smaller index are relatively higher in $P_{gw}$, the strategy is more inclined to observe the previous field immediately. Secondly, because it takes a relatively long time to slew the telescope to observe a new field, it also makes the algorithm tend to complete two times of observation in the same field and then perform a new field observation.
For Sequential Observation, the sequence is determined at the beginning, that is, we do the first time observation of all fields in order initially and then the second time observation.

Table \ref{tab:Time allocation for three typical events} displays the time allocation. Since the signal is strong enough and we do not need to spend too much time on a very small increase of detection probability, it could be found that the first observation of all fields in Sequential Observation is 1s. That means within a short exposure time, an expected signal-to-noise ratio can be achieved. For the same consideration, the first time observation for each field does not takes much time as well in Local Optimization. The second observation time of Sequential Observation is the same as the first observation, so it is fixed in 1s. The second observation time of Local Optimization shows great difference. If the difference of $P_{gw}$  between the former and later fields is not very large and the light curve changes rapidly( which means we will see the brightness differs obviously within a short second observation time), the second observation time may be only 1s. For the period when the difference of $P_{gw}$ between the former and later fields is relatively larger and the light change is relatively flatter, it will be required to continue the second time observation of the former field even until 220s.

\subsection{Multi-sample Events Comparison}
Then we will show the results of the simulated X-ray follow-up observations for the multi-sample events, using two algorithms as well as two quite different light curve model respectively.

\begin{figure}[ht]
\centering
\includegraphics[scale=0.5]{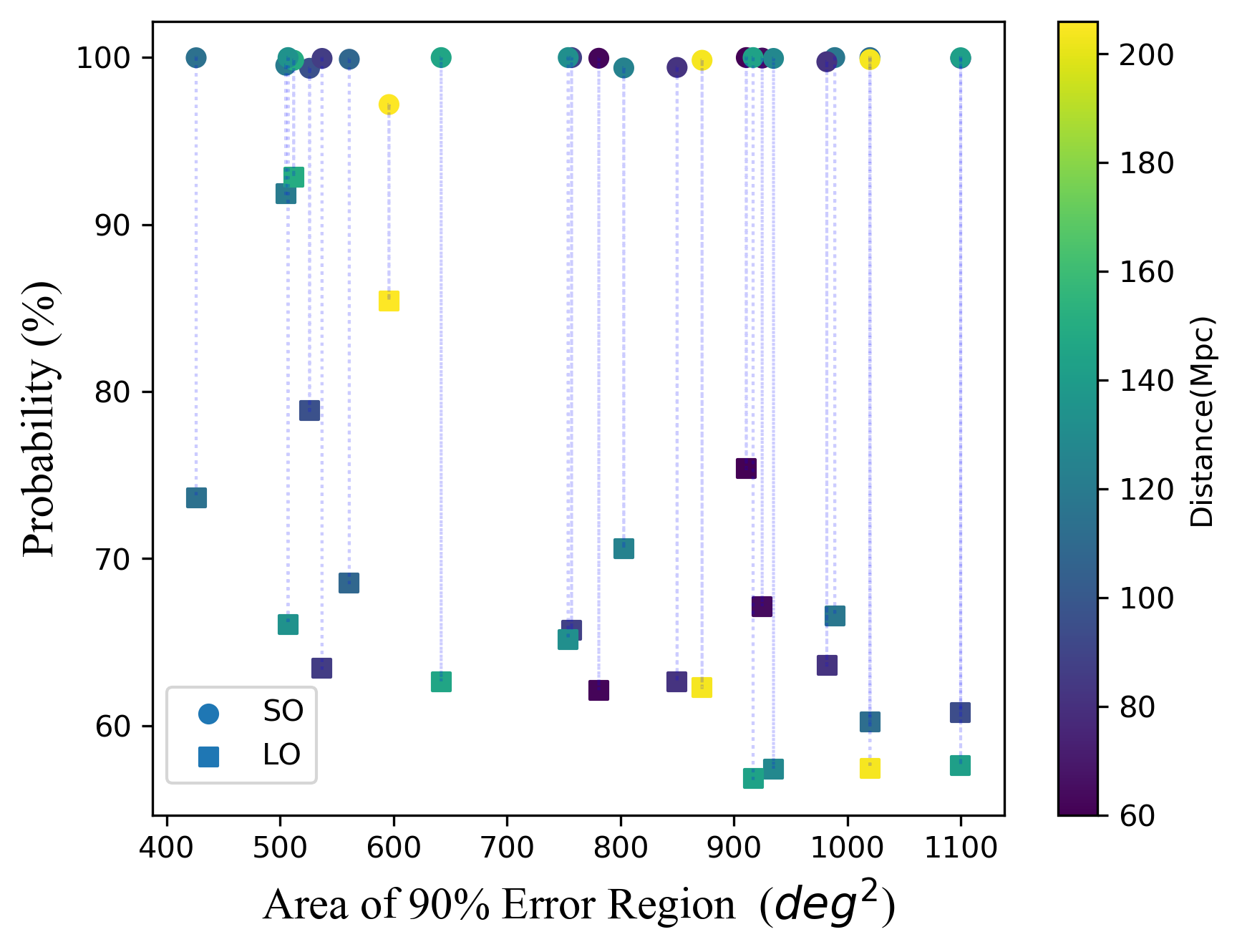}
\caption{The detection probability for the GRB afterglows model. The dashed line connects the results of different algorithms for the same event.}
\label{fig:X-ray detection probability for the GRB afterglows model}
\end{figure}

Firstly we can compare the efficiency of two algorithms in Figure \ref{fig:X-ray detection probability for the GRB afterglows model}. We use the GRB Afterglows Model here. Because the curve has a very bright signal in the early stage, if it can be observed soon after the outburst of GW events, it will be detected successfully without special strategic optimization. In addition, considering that the trigger of GW events and the telescope slewing to the expected position require time, so the initial observation time for strategy optimization of the GRB Afterglows Model is set at 1000s after the event outburst. The figure displays the optimized detection probability corresponding to different 90\% error region events, using two algorithms respectively. It can be seen that for the same event, compared with Local Optimization, Sequential Observation can have a significantly higher and more stable probability.

\begin{figure}[ht]
\centering
\includegraphics[scale=0.5]{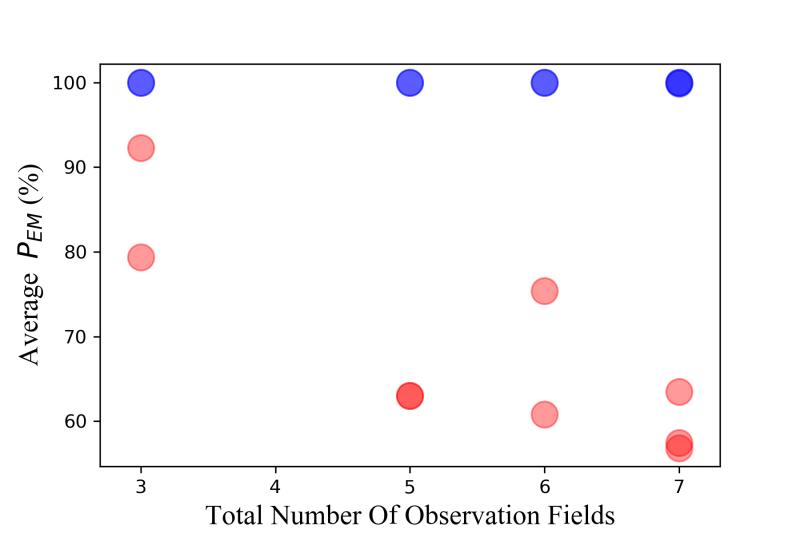}
\caption{The comparison of average $P_{em}$ corresponding to events with different total number of observation fields. Blue dots represent Sequential Observation and the red ones represent Local Optimization. Some dots are shown in darker color due to overlap of multiple events.}
\label{The comparison of average Pem corresponding to events with different total number of observation fields }
\end{figure}

There may be some slight fluctuations in the detection probability of Sequential Observation between different events. Because the probability of $P_{1}$ and $P_{2}$ in Sequential Observation is close to saturation, that is, both are close to 100\%, the fluctuation mainly results from the different $P_{gw}$ of the different events observation region. When the $P_{gw}$ of a new field is too small to observe, we will stop our first time observation and start the second time observation of all fields. There will be a maximum of about 3\% change, but it is basically stable above 95\%, which is also mentioned above. A trend can also be seen in the figure. For events with a smaller 90\% error region fields, the detection probability of Local Optimization is higher. Smaller 90\% error region fields lead to the smaller total observation fields to some extent(In fact, the total observation fields are also related to the size of the field of view and the shape of the 90\% error region, so we can see that the fields of the 90\% error region and the detection probability are not completely linearly related). With smaller observation fields, which means fewer observation fields as well, $P_{gw}$ is more concentrated on the first few observation fields. Therefore, when using Sequential Observation, more observation time can be spent on the first few observation fields with high $P_{gw}$, so that the detection probability in these fields is closer to saturation. For events with a large observation field, $P_{gw}$ is distributed in more observation fields, and the difference between the fields is not big enough to tilt the time allocation to fields with high $P_{gw}$ probability, resulting in the failure of the locally optimized observation strategy to concentrate time on a small part area of the high $P_{gw}$. So the larger observation area has a relative disadvantage in the total detection probability. In the Figure \ref{The comparison of average Pem corresponding to events with different total number of observation fields } we also can see that for the events with more observation fields,  $P_{em}$ may be smaller than those who have fewer observation fields, leading to the difference in final detection probability.
For Sequential Observation, since the second time observation is always made after a fairly long time interval, the probability of $P_{2}$ can always be nearly saturated. So it is only necessary to ensure that the signal is large enough when we are doing the first time observation. Then the probability of the first time observation will be close to saturation, and we can get a good overall observation probability.
\begin{figure}[ht]
\centering
\includegraphics[scale=0.5]{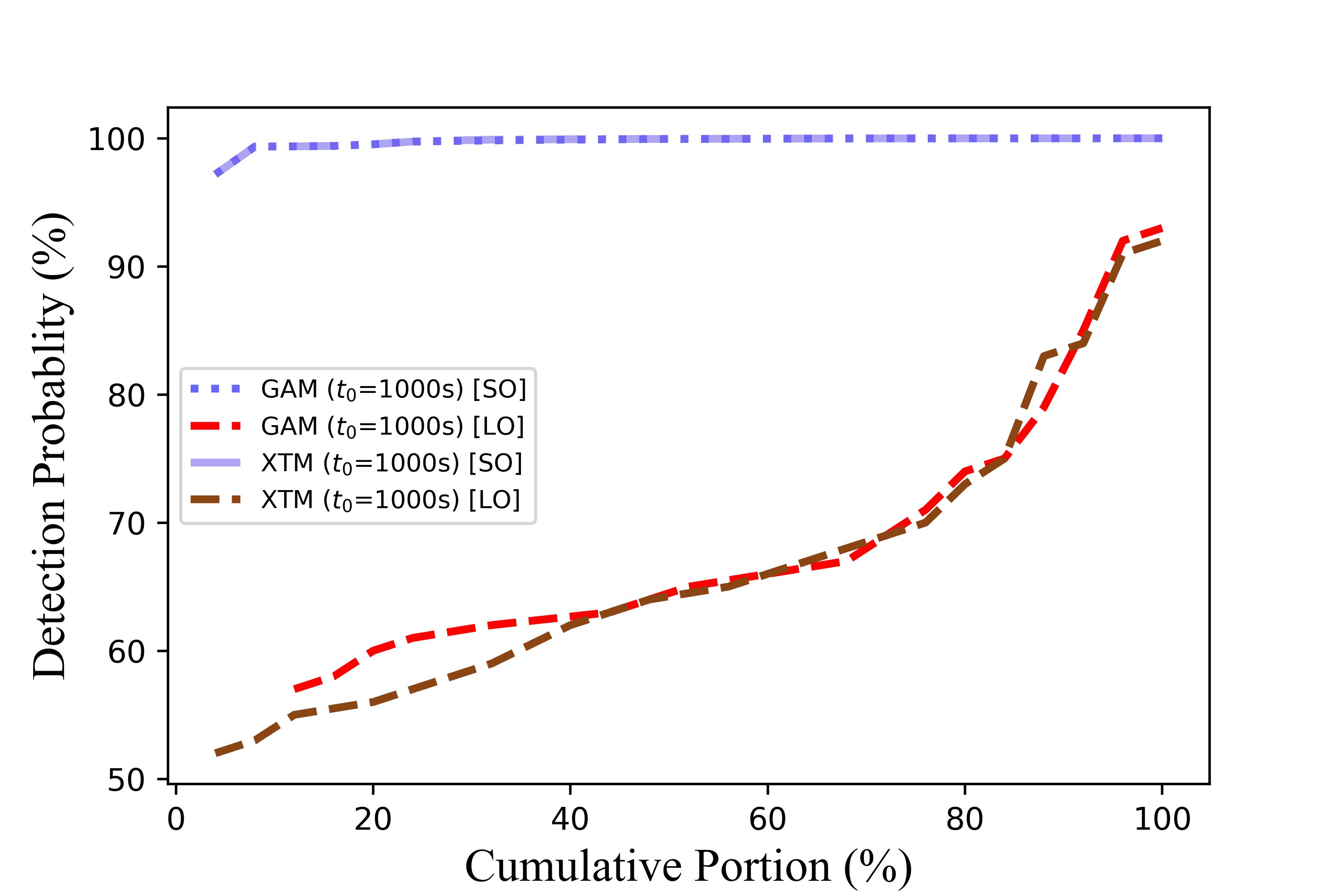}
\caption{Cumulative distribution of detection probability. Different lines represent the results of two different models using different algorithms ($t_{0}$ refers to the start time)}.
\label{fig:CDF1}
\end{figure}


We work our another light curve, the X-ray Transient Model, with our calculations as well. Figure \ref{fig:CDF1} display the results. We still get a conclusion basically similar to the GRB Afterglows Model. It shows that under different models, Sequential Observation has shown significantly better results than Local Optimization.

We further consider that applying the same strategy, which means the same observation time and observation fields sequence, obtained in the GRB Afterglows Model to the the X-ray Transient Model. The Sequential Observation still shows its superiority. The detection probability is close to 100\%, that is, the total probability is still basically determined by $P_{gw}$ of observed sky area as long as the source is bright enough to finish the first time observation. Local Optimization shows relatively poor results in both different models.

We also compare the results of different models with the same strategy. The detection probability results of Sequential Observation under different models are basically the same. While using the same strategy given by Local Optimization in a different model, we get generally lower detection probability. This result actually illustrates the strategy based on the local optimization of the detection probability change in the shortest step, depends heavily on the model in the actual calculation. However, Sequential Observation, the strategy of doing the first time observation of all fields initially, can better ensure that the time interval between two observations of the same field is long enough, and the probability of $P_{2}$ can be close to saturation. So as long as the signal is not too weak to observe, even for different models, Sequential Observation will give a good detection probability. Therefore, Sequential Observation is not highly dependent on the model.

\section{Discussion and conclusion}\label{sec:conclusion}

In this work, we devised and compared two different algorithms to optimize the probability of a successful X-ray afterglow detection triggered by GW alerts.
We apply the EP's WXT as a candidate telescope, which has a large FOV and can get most of the $P_{gw}$ with handful fields.
Interestingly, the simpler strategy, or Sequential Observation, which is to finish the first time observation of as many fields as possible initially, constantly outperform the more complicated algorithm. 
The results indicates that multiple observations of the same fields in fairly long time interval would have better chance to observe the afterglow. 
We notice that large FOV will be very beneficial for rapid detection of afterglow.
Indeed, for the earlier cases, the afterglow is expected to be bright enough that a very short exposure time is sufficient for detection, the majority of time is actually spent on the slew of telescopes.

There are multiple aspects we can further explore.
For example, our threshold for distinguishing the flux change using only the information of integrated photon numbers, while X-ray telescopes can register the arrival time of X-ray photons, which can further help distinguish random fluctuation from actual change of flux.
This analysis based on the assumptions of a static telescope with unconstrained pointing and the independence of statistical uncertainty between the distance and the GW trigger sky location.
Space-borne gravitational wave detectors like TianQin has the potential of predicting the merger with very high accuracy \cite{Liu:2020},  and a co-ordinated observation can better depict the very early stage evolution.
All issues can help shape a more realistic and more promising future of successful multi-messenger astronomy.




To extend this work, it may be helpful to consider the further constraints that arise from the diurnal cycle, observing time available for afterglows, limitations on the pointing a particular telescope is capable of, and the rise and set of tiles.

\acknowledgments
{We would like to thank Binbin Zhang, Jian-dong Zhang, Yuanpei Yang, Ruiqing Liu, Michael Coughlin and Mouza Mualla for helpful comments and discussions.
This work was supported in part by the National Natural Science Foundation of China (Grants No. 11703098) and Guangdong Major Project of Basic and Applied Basic Research (Contract No. 2019B030302001).
H.S. acknowledge support by the Strategic Pioneer Program on Space Science, Chinese Academy of Sciences, grant No. XDA15052100, XDA15310300 and the Strategic Priority Research Program of the Chinese Academy of Sciences grant No. XDB23040100.
This work made use of data supplied by the UK Swift Science Data Centre at the University of Leicester.
}

\bibliographystyle{apsrev4-1}
\bibliography{reference}

\end{document}